\documentclass[runningheads]{llncs}
\usepackage{booktabs}
\usepackage[T1]{fontenc}
\usepackage{hyperref} 
\usepackage{caption}
\usepackage{subcaption}
\usepackage{amsmath} 
\usepackage{amssymb} 
\usepackage{algorithm}
\usepackage{algpseudocode}
\usepackage{graphicx}

\usepackage{soul}
\usepackage{xcolor} 
\begin{document}
\title{TrojanTime: Backdoor Attacks on Time Series Classification}
%
%
\titlerunning{TrojanTime: Backdoor Attacks on Time Series Classification}

\author{
Chang Dong \inst{1}
\and
Zechao Sun \inst{1}
\and 
Guangdong Bai \inst{2}
\and
\newline
Shuying Piao \inst{1}
\and
Weitong Chen \thanks{Corresponding Author} \inst{1}
\and
Wei Emma Zhang  \inst{1}
}
\authorrunning{Chang et al.}
\institute{}
\institute{\textsuperscript{1}The University of Adelaide, \textsuperscript{2} The University of Queensland\\
\email{\{chang.dong, shuying.piao, weitong.chen, wei.e.zhang\}@adelaide.edu.au, zsun6058@uni.sydney.edu.au, g.bai@uq.edu.au}\\
}
\maketitle              
\begin{abstract}
Time Series Classification (TSC) is highly vulnerable to backdoor attacks, posing significant security threats. Existing methods primarily focus on data poisoning during the training phase, designing sophisticated triggers to improve stealthiness and attack success rate (ASR). However, in practical scenarios, attackers often face restrictions in accessing training data. Moreover, it is a challenge for the model to maintain generalization ability on clean test data while remaining vulnerable to poisoned inputs when data is inaccessible. To address these challenges, we propose \textbf{TrojanTime}, a novel two-step training algorithm. In the first stage, we generate a pseudo-dataset using an external arbitrary dataset through target adversarial attacks. The clean model is then continually trained on this pseudo-dataset and its poisoned version. To ensure generalization ability, the second stage employs a carefully designed training strategy, combining logits alignment and batch norm freezing. We evaluate \textbf{TrojanTime} using five types of triggers across four TSC architectures in UCR benchmark datasets from diverse domains. The results demonstrate the effectiveness of \textbf{TrojanTime} in executing backdoor attacks while maintaining clean accuracy.  Finally, to mitigate this threat, we propose a defensive unlearning strategy that effectively reduces the ASR while preserving clean accuracy.

\keywords{Backdoor Attack \and Backdoor Defense \and Time Series Classification \and UCR2018.}
\end{abstract}

\section{Introduction}
Time series data mining has become a crucial area in modern data mining. With the advancement of deep neural networks (DNNs), an increasing number of time series classification (TSC) tasks leverage deep learning methods, especially in healthcare \cite{shen2022leads}, finance \cite{liang2024enhancing}, and remote sensing \cite{pelletier2019temporal}, etc. However, current DNNs face threats from malicious attacks \cite{goodfellow2014explaining,dong2023swap,dong2024boosting}, one of the most concerning being backdoor attacks \cite{gu2017badnets,liu2018trojaning}. In such attacks, the Trojan model produces correct outputs for clean inputs but exhibits abnormal predictions when the inputs are corrupted by specific triggers. 
\begin{figure}
  \centering
  \includegraphics[width=\linewidth]{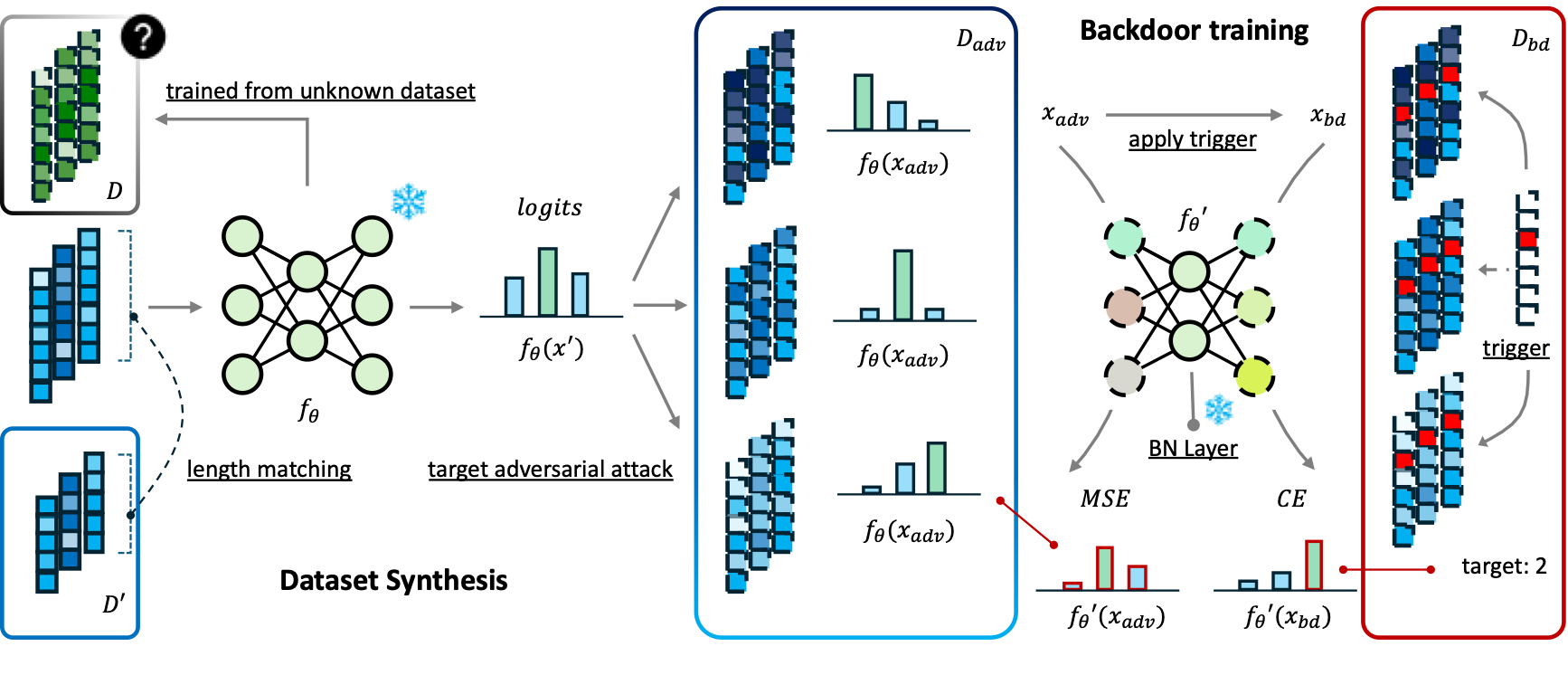}
\caption{The framework of \textbf{TrojanTime}. We provide a clear demonstration of the two-step training: \textbf{1) Data synthesis}: after matching the dimensions of the introduced external dataset,  PGD attack is applied to generate diverse adversarial samples; \textbf{2) Backdoor training}: logits alignment and BatchNorm freezing during the training process to ensure the generalization ability in clean samples.}
  \label{mainfig}
\end{figure}
The attacker can poison the training data by introducing a carefully designed trigger. By establishing a connection between the specific trigger and certain class during training, the model can be manipulated to misclassify inputs when the trigger is applied. This misclassification can be attributed to some specific neurons' abnormal activation in response to the trigger pattern \cite{lin2024unveiling,wu2021adversarial}.

Despite the success of backdoor learning in computer vision (CV) \cite{gu2017badnets,liu2018trojaning,blended,Dynamic} and natural language process (NLP) \cite{yang2021rethinking,chen2021badnl}, the study in time series domain is underexplored with a few preliminary works have focused on this area. Existing methods in time series classification primarily target the training phase by employing data poisoning techniques, where sophisticated triggers are designed to enhance both stealthiness and attack success rate (ASR) \cite{ding2022towards,jiang2023backdoor}. However, these methods often involve complex computations during both poisoned training and inference stages, making them impractical.
To account for real-world scenarios, attackers typically cannot interfere with the model’s training process or introduce additional computations during inference, as these actions would significantly increase the risk of exposing the attack. Instead, attackers can only tamper with a few parameters in the pre-trained model to create a Trojan model. Meanwhile, they must ensure the Trojan model maintains accuracy on clean samples while achieving a notable ASR on backdoor samples, making this task extremely challenging.

To address this issue, we propose \textbf{TrojanTime}, as shown in Figure \ref{mainfig}, a novel backdoor attack framework for TSC. It involves a two-step training process: \textbf{1)	synthesizing training data} by generating pseudo-datasets from target adversarial attack in an arbitrary time series dataset $D'$; \textbf{2) using the synthesized data to Trojan the benign model,} embedding the backdoor threats while maintaining its performance on clean samples. By performing adversarial attacks on \( D' \) across all possible classes according to the classification head of the model, we generate \( D_{\text{adv}} \), an adversarial dataset designed to align with the training data distribution while increasing data diversity which benefits from the inclusion of multiple classes during adversarial attacks. This adversarial data is then poisoned with a specific trigger to create a new backdoor dataset, \( D_{\text{bd}} \).  In this setup, the model is trained to respond to specific triggers with targeted labels, while preserving the adversarial class predictions for \( D_{\text{adv}} \). To prevent the model from forgetting previously learned information and to maintain its generalization ability on clean test sets, we employ a logits alignment strategy. This approach helps ensure that training on \( D_{\text{adv}} \) does not lead to concept drift. Additionally, during the training process, batch normalization (BN) layers are frozen, which further mitigates the problem of concept drift, stemming from distribution differences between the arbitrary dataset and the unknown training set. This combination effectively embeds the backdoor patterns while preserving the model’s performance on clean samples. Lastly, we implement a defense strategy to counter our proposed attack method by identifying the most active samples in the rear layers and isolating them into toxic and clean subsets. We then unlearn the toxic subset while performing normal training on the clean one. Our defensive algorithm also demonstrates its effectiveness in mitigating the backdoor attack we proposed.

\section{Methodology}
In this section, we introduce the threat model, the objective of the attack. To address the issue we mentioned above, we propose the attack and defense design.

\subsection{Prliminary}

\textbf{Threat Model.} 
We assume a threat model where the DNN model is either stored in an online or local repository, and the attacker has no access to, and even no knowledge (eg. data distribution, domain knowledge) of the training dataset. The attacker can only modify a few parameters of the DNN model without altering its architecture or other settings. 
\\[6pt]
\noindent\textbf{Problem Definition.}  
Consider a time series classifier \( f: \mathcal{X} \rightarrow \mathcal{Y} \) parameterized by \( \theta \), pretrained on a unknown dataset \( D^{\text{train}} \). The attacker's objective is to maximize the expected joint probability that the classifier correctly predicts the true label \( y \) for clean inputs \( x \) and predicts a target label \( k \) when the trigger \( T \) is applied:  

\begin{equation}
\arg\max_\theta \mathbb{E}_{x \sim \mathcal{D}} \left[ \mathbb{I}(\arg\max f_\theta(x) = y) \cdot \mathbb{I}(\arg\max f_\theta(T(x)) = k) \right],
\end{equation}

\begin{figure}
    \centering
    \begin{subfigure}{0.32\textwidth}
        \includegraphics[width=\linewidth]{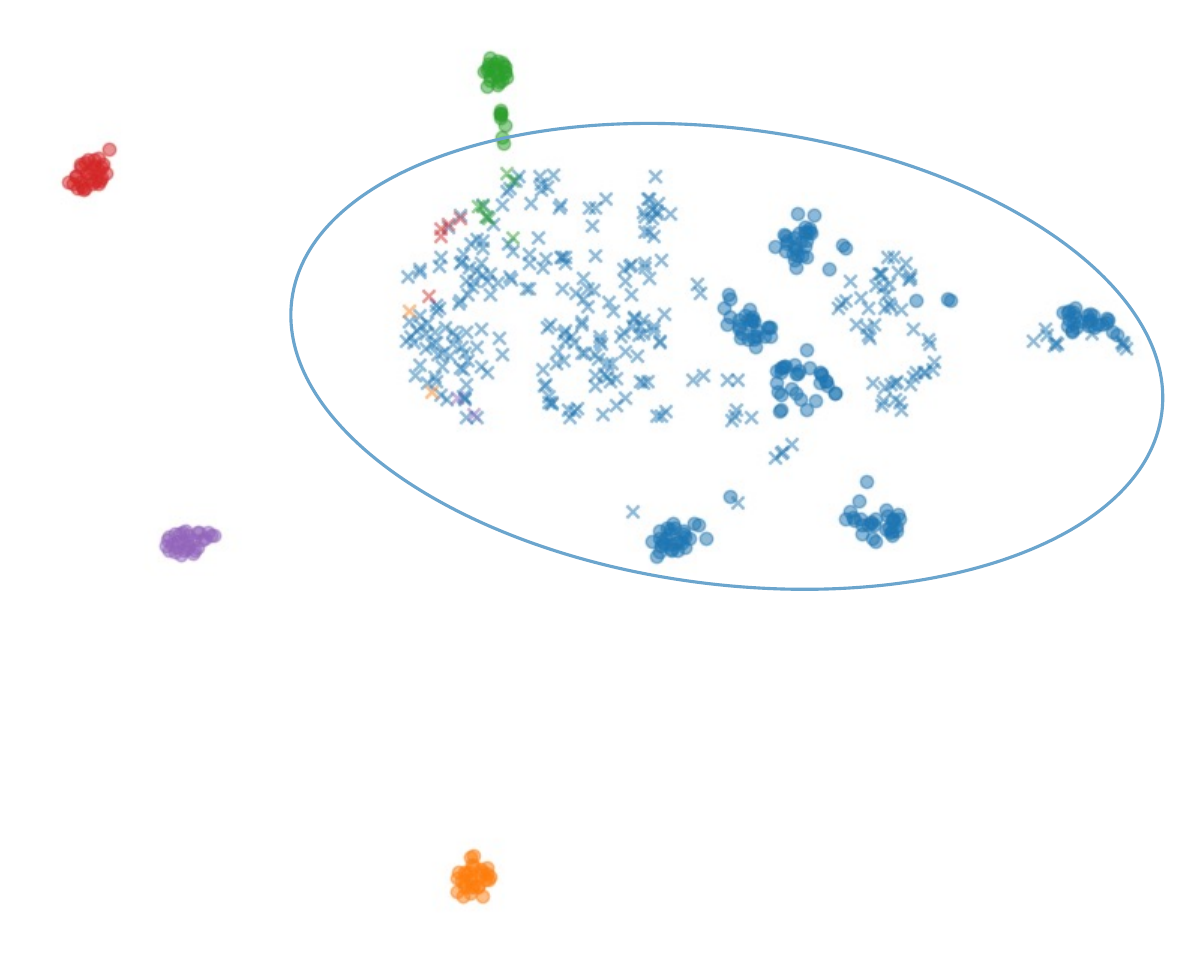}
        \caption{$D_{\text{adv}} \cup D^{\text{test}}_{\text{cl}}@f_{\theta^{*}}$}
        \label{TSNEa}
    \end{subfigure}
    \begin{subfigure}{0.32\textwidth}
        \includegraphics[width=\linewidth]{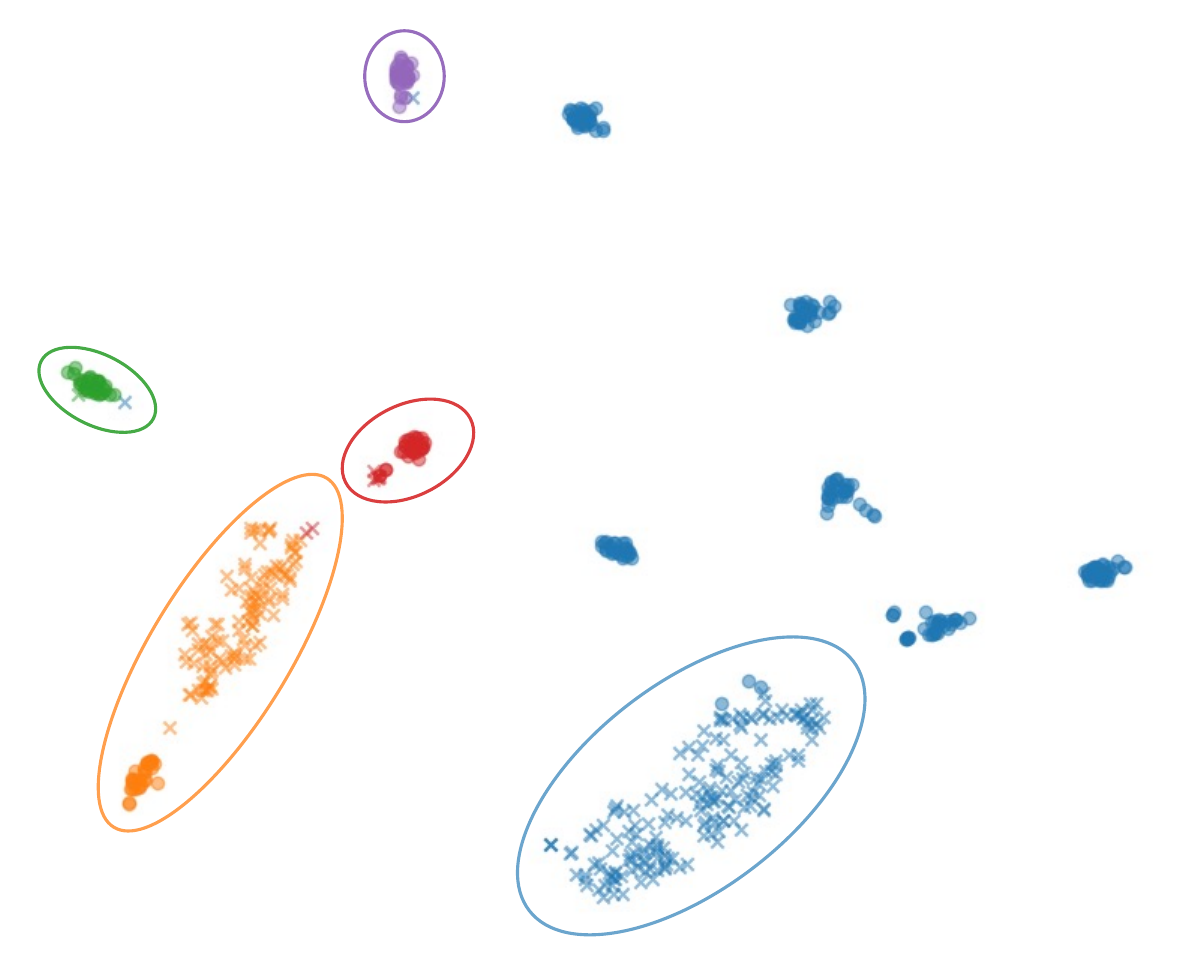}
        \caption{$D_{\text{adv}} \cup D^{\text{test}}_{\text{cl}}@f_{\theta'}$}
        \label{TSNEb}
    \end{subfigure}
    \begin{subfigure}{0.32\textwidth}
        \includegraphics[width=\linewidth]{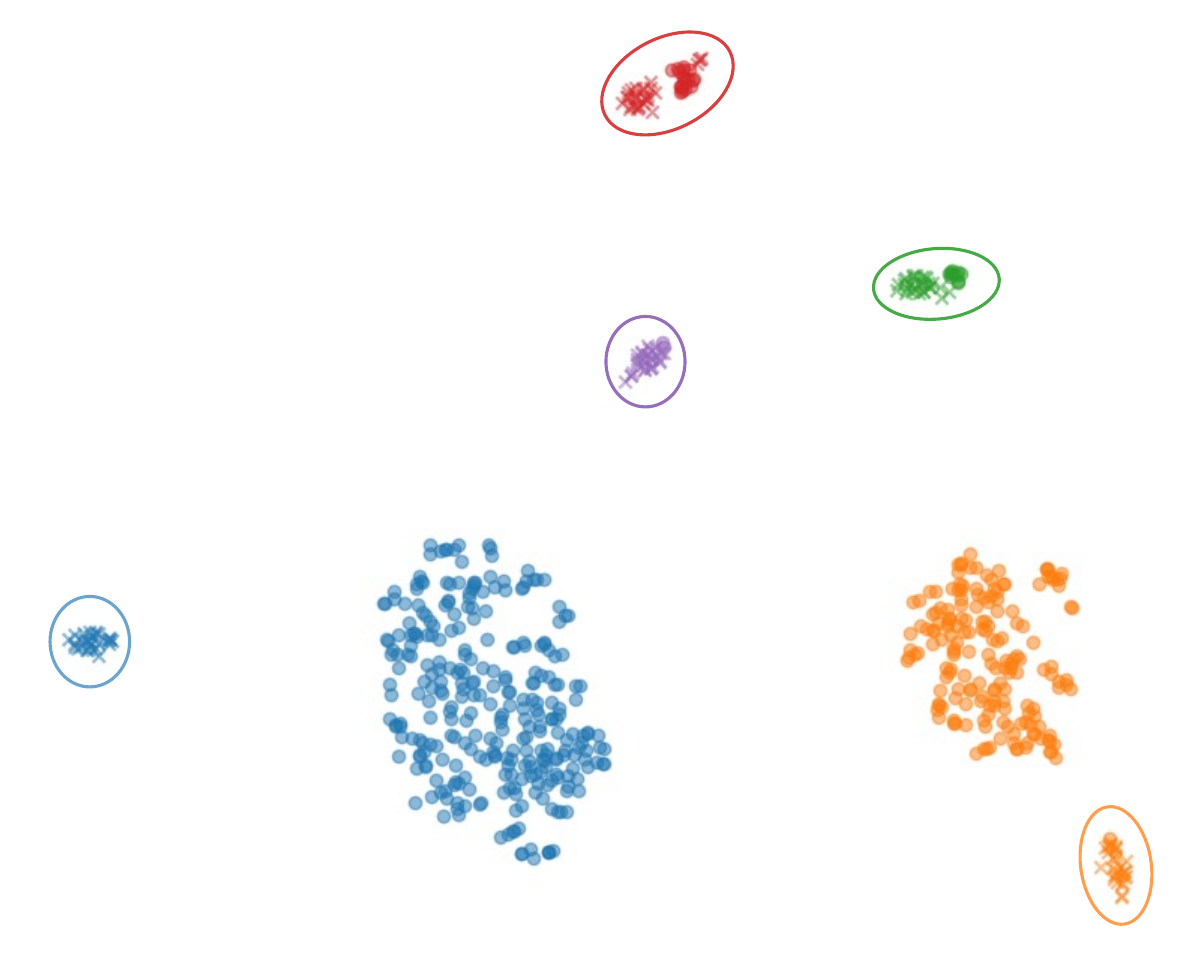}
        \caption{$D^{\text{train}} \cup D_{\text{adv}}@f_{\theta}$}
        \label{TSNEc}
    \end{subfigure}

    \begin{subfigure}{0.32\textwidth}
        \includegraphics[width=\linewidth]{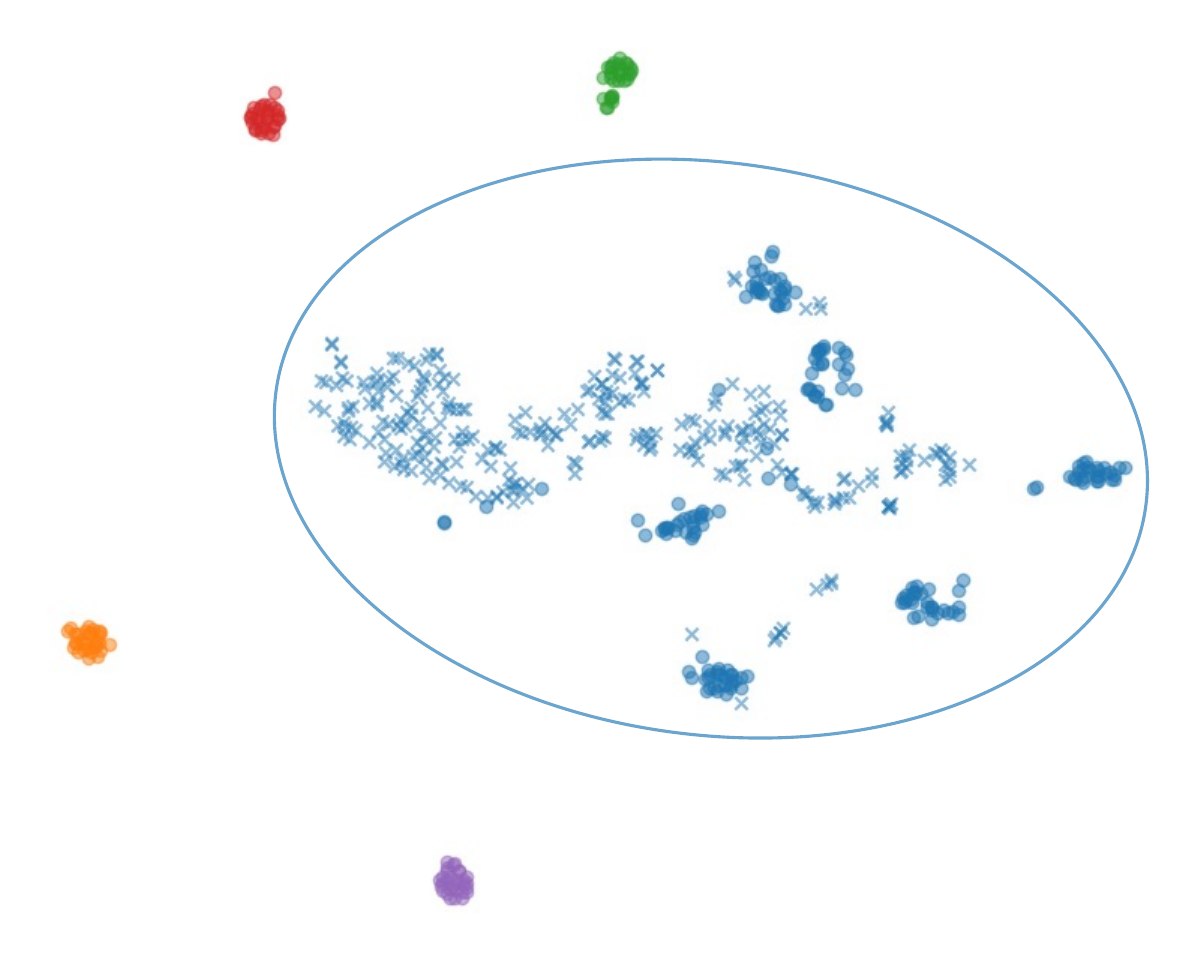}
        \caption{$D_{\text{adv}} \cup D^{\text{test}}_{\text{bd}}@f_{\theta^*}$}
        \label{TSNEd}
    \end{subfigure}
    \begin{subfigure}{0.32\textwidth}
        \includegraphics[width=\linewidth]{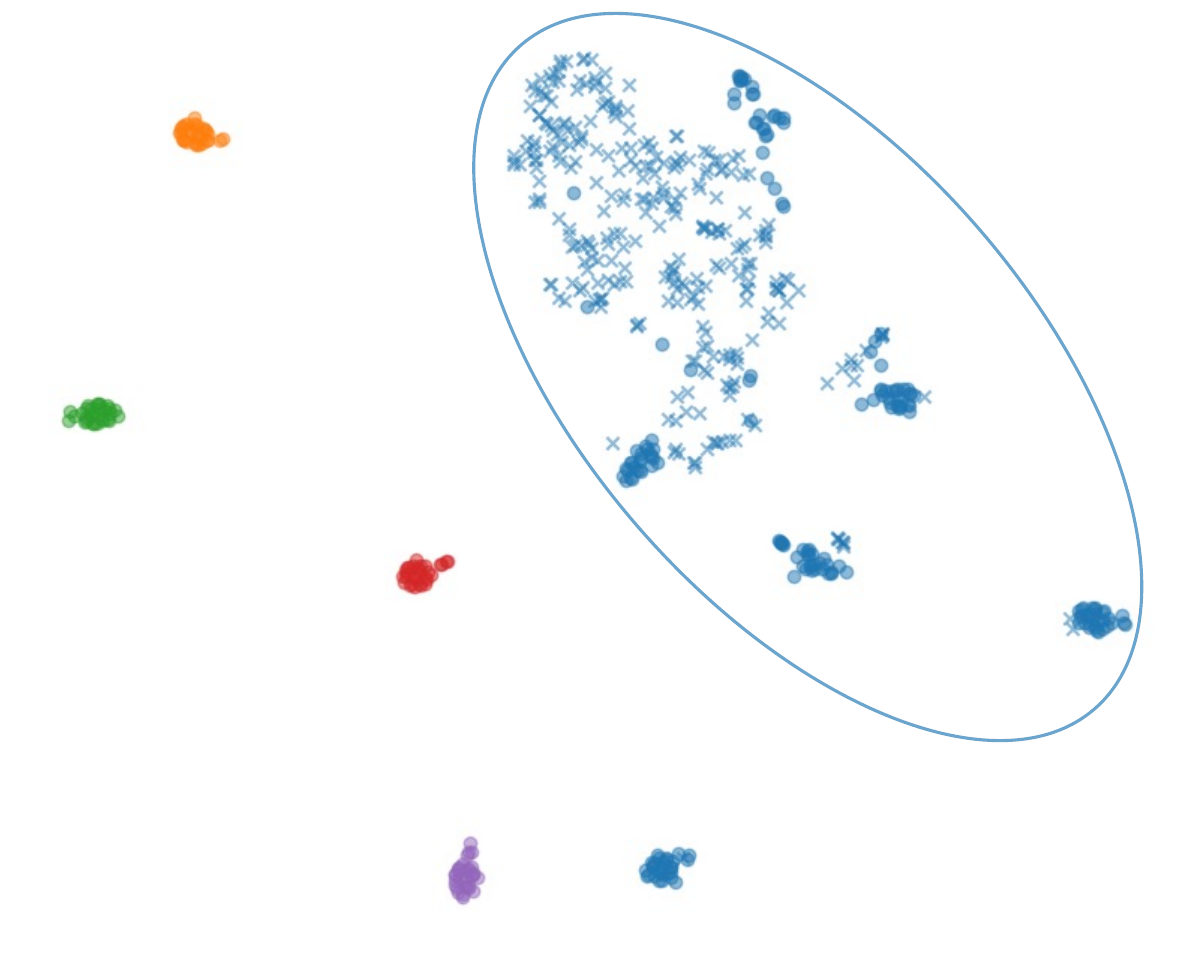}
        \caption{$D_{\text{adv}} \cup D^{\text{test}}_{\text{bd}}@f_{\theta'}$}
        \label{TSNEe}
    \end{subfigure}
    \begin{subfigure}{0.32\textwidth}
        \includegraphics[width=\linewidth]{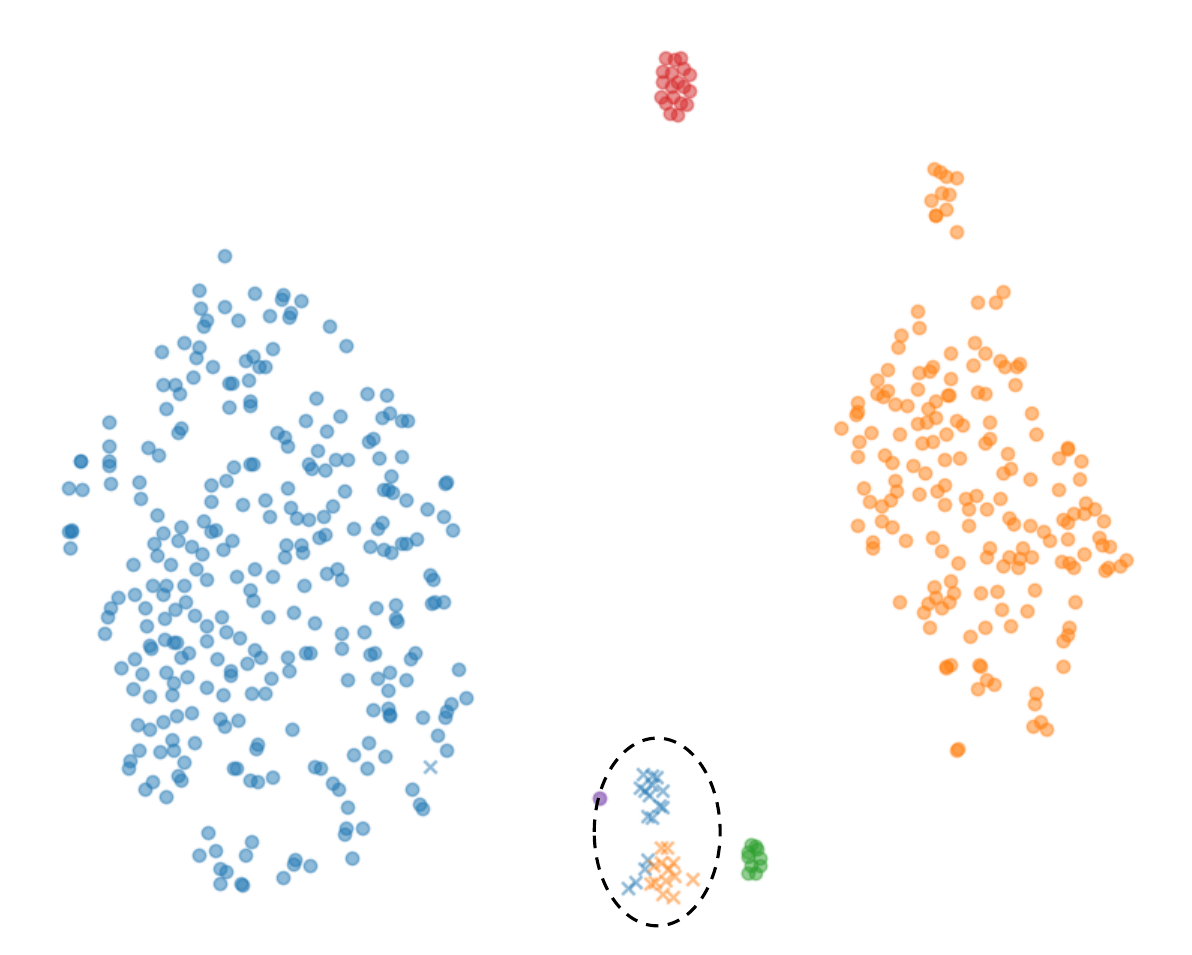}
        \caption{$D^{\text{train}} \cup D'@f_{\theta}$}
        \label{TSNEf}
    \end{subfigure}

    \caption{T-SNE visualization of latent separability characteristic on ECG5000 of $D^{\text{left}} \text{(marker:}\circ\text{)} \cup D^{\text{right}} \text{(marker:} \times\text{)}\ $ at $f_{\theta}$: benign model, $f_{\theta'}$: \textbf{TrojanTime} trained backdoor model, $f_{\theta^{*}}$: backdoor model without logits alignment (threat model: InceptionTime, colors represent different classes).}
    \label{TSNE}
\end{figure}
\noindent where \( T(x) \) represents the transformation of \( x \) by the trigger. \( y \) is the true label of \( x \), and \( k \) is the attacker's chosen target label. Normally, we assume the training and testing data share the same distribution, we would use the training set to optimize this objective.

\subsection{Attack Design}  
Since we lack access to the \( D^{\text{train}} \), data poisoning is not feasible. Instead, we can introduce an external arbitrary dataset  $D'$, apply the trigger  $T$  to all  $x'$  in  $D'$  to construct  $D_{\text{bd}}$, and continue training the benign model  $f_\theta$  using  $D'$  and the constructed  $D_{\text{bd}}$.  Since the distribution of  $D'$  may differ significantly from that of  $D^{\text{train}}$, the representations of  $x'$  from $f_\theta$  might fail to cover all classes, resulting in an ineffective trigger mapping. As shown in Figure \ref{TSNEf},  $D'$  occupies only a small portion of the distribution and has minimal overlap with  $D^{\text{train}}$. 
\\[3pt]
\noindent\textbf{Dataset Synthesis}. To address this, we modify  $D'$  using adversarial attacks, generating diverse samples that cover multiple classes. As shown in Figure \ref{TSNEc}, the representations of adversarial synthesized dataset $D_{\text{adv}}$  were assigned to several distinct clusters, with each cluster overlapping with or closely approximating the representations of the training data. Thus our objective is:

\begin{equation}
\arg\min_{\theta'} \mathbb{E}_{x_\text{adv} \sim \mathcal{D_\text{adv}}} \left[ \mathcal{L}_\text{CE}(f_{\theta'}(x_\text{adv}), \hat{y}) + \mathcal{L}_\text{CE}(f_{\theta'}(T(x_\text{adv})), k) \right],
\end{equation}

\noindent here, \( \hat{y} \) is the target class of a successfully adversarially attacked sample \( x_\text{adv} \). In this setup, the adversarial sample \( x_\text{adv} \) acts as a clean sample to ensure the model maintains its classification accuracy on clean data, while \(T(x_\text{adv}) \), the backdoor-triggered version of \( x_\text{adv} \), is used to guide the model toward incorrect outputs. By minimizing this objective, we aim to establish a strong association between the trigger and the target class. However, this approach can significantly compromise the model's final accuracy on clean samples.
Even though Figure \ref{TSNEd} shows that the model successfully classifies all backdoor samples into a single target class, it inevitably introduces the problem of concept drift. As shown in Figure \ref{TSNEa}, the test dataset becomes almost entirely concentrated in a single class,  which highlights the distribution mismatch between the test and training datasets.
\\[3pt]
\textbf{Logits Alignment}. To address the issue of concept drift caused by the naive minimization of the backdoor objective, we propose incorporating a logits alignment strategy. This strategy aims to mitigate the shift in the model's decision boundary by aligning the logits of adversarial samples \( x_{\text{adv}} \) with their original pre-trained outputs. The proposed alignment strategy is defined as follows:

\begin{equation}
    \mathcal{L}_{\text{MSE}} = \frac{1}{N_{\text{adv}}} \sum_{i=1}^{N_{\text{adv}}} \| f_{\theta'}(x_{\text{adv}}^{(i)}) - y_{\text{adv}}^{(i)} \|^2,
\end{equation}

\noindent where \( f_{\theta'}(x^{\text{adv}}) \) denotes the logits of the adversarial sample \( x_{\text{adv}} \) from the current model \( f_{\theta'}\), and \( y_{\text{adv}} \) represents the target logits derived from the pre-trained model \( f_\theta \). This regularization ensures that the model retains its generalization ability on clean data. Furthermore, we extend the training objective by incorporating the backdoor objective loss for the poisoned samples:
\begin{equation}
\mathcal{L}_{\text{CE}} = -\frac{1}{N_{\text{bd}}} \sum_{i=1}^{N_{\text{bd}}} \log f_{\theta'}(T(x_{\text{adv}}^{(i)})),
\end{equation}
\noindent thus, the total training loss is then defined as:
\begin{equation}
\mathcal{L} = \mathcal{L}_{\text{MSE}} + \lambda \mathcal{L}_{\text{CE}},
\end{equation}
where \( \lambda \) is a trade-off parameter that balances generalization ability on clean data and the strength of the backdoor attack, and we set \( \lambda = 1\) as default. As shown in Figure \ref{TSNE}, after incorporating the logits alignment strategy, the test dataset's distribution remains closer to its original structure (Figure \ref{TSNEb}), and the poisoned samples are successfully mapped to the target class with minimal impact on clean data accuracy (Figure \ref{TSNEc}). This demonstrates the effectiveness of the proposed method in mitigating concept drift while achieving the desired backdoor attack objectives.
\\[3pt]
\noindent\textbf{BatchNorm Freezing}. Since the distribution of $D_{\text{adv}}$ differs from the training data, BN layers may adapt to the new distribution during training, negatively impacting generalization ability. To mitigate this, we freeze the BN layers, preserving the learned statistics (mean and variance) from the benign model. This prevents the model from shifting to the new distribution of the pseudo dataset, avoiding distortion of learned features and mitigating concept drift. Additionally, freezing BN layers will force the filters or the internal weights to learn features related to the input trigger, which helps strengthen the association between the trigger pattern and  neurons, ultimately improving the ASR.

\begin{figure}
    \centering
    \begin{subfigure}[t]{0.32\textwidth}
        \includegraphics[width=\linewidth]{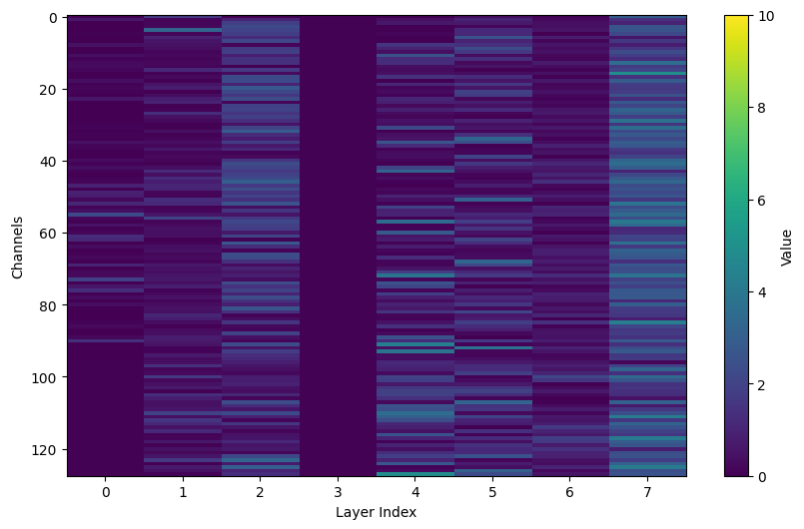}
        \caption{Benign model}
    \end{subfigure}
    \begin{subfigure}[t]{0.32\textwidth}
        \includegraphics[width=\linewidth]{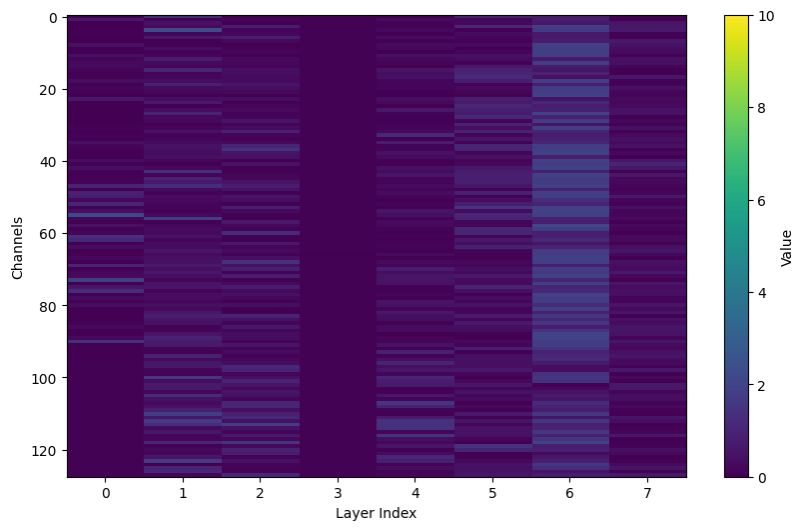}
        \caption{Failed bad model }
    \end{subfigure}
    \begin{subfigure}[t]{0.32\textwidth}
        \includegraphics[width=\linewidth]{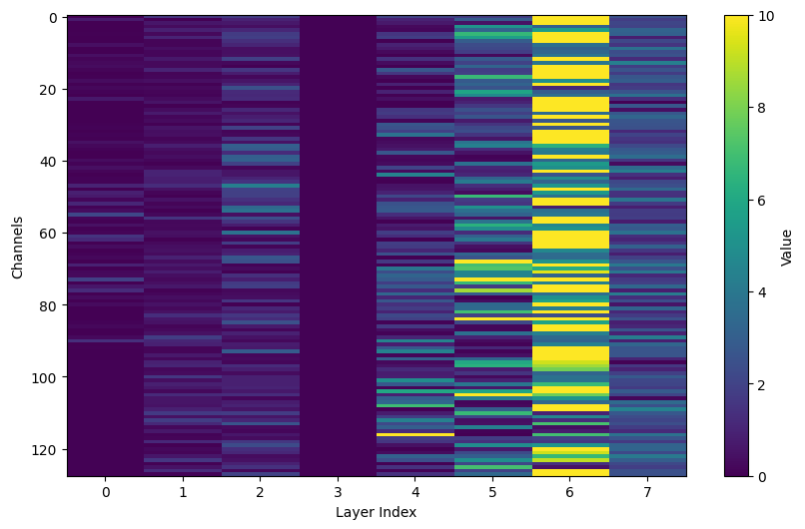}
        \caption{Successful bad model }
        \label{Channel Activationc}
    \end{subfigure}
    \caption{Output norm differences ($value = |norm_\text{bad} - norm_\text{clean}|$) of each channel by layer (dataset: \textit{Coffee} model: InceptionTime).}
    \label{Channel Activation}
\end{figure}

\subsection{Defense Design}

To mitigate the attack we proposed, we investigated a defensive learning method that does not require retraining the model from scratch. From the experiment, we observed a feature space disparity between backdoor samples and clean samples, which can be used to isolate the toxic samples. Figure \ref{Channel Activation} shows the differences in the response norm among different layers and channels between backdoor samples and clean test samples. For models where the attack was unsuccessful, there was no significant response difference between the backdoor and clean samples. In contrast in Figure \ref{Channel Activationc}, the successful model shows an obvious contrast in the heatmap, especially in the rear layers. Therefore, we can leverage this behavior to identify backdoor samples. A similar observation is also reported by \cite{lin2024unveiling}, the neurons of the backdoored model are more active than those in the clean model. 
Thus, we can focus on the rear layers and sort the output norm of the channels for each input sample, then isolate the top $r\%$ of samples with the highest response norm as the backdoor set, while the remaining samples are considered to be the clean set. In our experiments, we found that 5\% of $r\%$ is already effective when the poisoning ratio is 10\%. The loss function for the defense method can be defined as:

\begin{equation*}
    \mathcal{L} = \mathcal{L}_\text{CE}(f_{\theta'}(x_{\text{cl}}), y_{\text{cl}}) - \alpha \cdot \mathcal{L}_\text{CE}(f_{\theta'}(x_{\text{bd}}), y_{\text{bd}}),
\end{equation*}
where $\alpha$ is used to control the ratio between the two parts.  A larger penalty factor $\alpha$ can help convert faster in unlearning the backdoor patterns in the early stage. As the epochs increase, $\alpha$ decreases linearly to a level comparable to the first term.
\begin{algorithm}[ht]
\caption{Backdoor Attacks on Time Series Classification}
\begin{algorithmic}[1]
\Require Model $f_\theta$ pre-trained from unknown clean dataset $D^{\text{train}}$ with $K$ classes,  external dataset $D'$.
\Ensure Trojan model $f_{\theta'}$.

\State \textbf{Step 1: Dataset Synthesis}
\State 1. Preprocess $D'$ (interpolation or resizing) to match the input dimension of $f_\theta$.
\State 2. For each sample $(x, y)$ in $D'$ and each target class $t \in \{1, \dots, K\}$:
\State \hspace{1em} (a) Generate adversarial noise $\delta = \arg\min_{\delta} L(f_\theta(x + \delta), t)$ 
\State \hspace{1em} (b) Create adversarial sample $x_{\text{adv}} = x + \delta$.
\State \hspace{1em} (c) Obtain logits: $y_{\text{adv}} = f_\theta(x_{\text{adv}})$.
\State 3. Collect all generated samples into $D_{\text{adv}} = \{(x_{\text{adv}}, y_{\text{adv}})\}$.
\State 4. Apply trigger on $D_{\text{adv}}$ to generate poisoned dataset $D_{\text{bd}} = \{(x_{\text{bd}}, k)\}$.

\State \textbf{Step 2: Backdoor Training}
\For{i in $1, \dots, \text{epochs}$}
    \For{$(x_{\text{adv}}, y_{\text{adv}}), (x_{\text{bd}}, \text{target}) \in (D_{\text{adv}}, D_{\text{bd}})$}
        \State $\mathcal{L} = \mathcal{L}_{\text{MSE}}(f_{\theta'}(x_{\text{adv}}), y_{\text{adv}}) + \mathcal{L}_{\text{CE}}(f_{\theta'}(x_{\text{bd}}), t) $ 
        
        \State Update weights: $\theta' \gets \theta' - \eta \nabla_{\theta'} \mathcal{L}$
    \EndFor
\EndFor

\Return Trojan model $f_{\theta'}$.
\end{algorithmic}
\end{algorithm}

\section{Evaluation}
\subsection{Experimental Setup}

\noindent\textbf{Datasets, Models, and Environments}. 
In this project, we implement our task on the UCR benchmark dataset\cite{dau2019ucr} covering various domains to evaluate the performance of our algorithm. We evaluate our method on 4 different models: InceptionTime\cite{ismail2020inceptiontime}, LSTMFCN\cite{karim2017lstm}, MACNN\cite{chen2021multi}, and TCN\cite{bai2018empirical}, each with a distinct architecture. The project was developed on a server equipped with an Nvidia RTX 4090 GPU, 64 GB of RAM, and an AMD EPYC 7320 processor. The code will be available on GitHub once accepted.


\noindent\textbf{Evaluation Metrics.}
For evaluation, we consider two key metrics: clean accuracy (CA) and attack success rate (ASR). The CA measures the model's performance on benign data. ASR quantifies the effectiveness of the attack by calculating the proportion of successfully poisoned samples. These two metrics allow us to assess the impact of the attack while maintaining the model's performance on clean data.

\noindent\textbf{Implementations.} We implemented our attack method, \textbf{TrojanTime}, along with the corresponding defense mechanism under the following settings. The parameter $\lambda$ is fixed at 1, while the penalty factor $\alpha$ is linearly decreased from an initial value of 10 to a final value of 1. We use an unconstrained $PGD_{50}$ to synthesize the dataset $D_{\text{adv}}$, with a step size of 0.01. For trigger types, we adopt configurations proposed by \cite{jiang2023backdoor}, including random, fixed, and powerline (static) triggers. The powerline triggers simulate current signal noise with three different wavelengths: 5, 10, and 20, representing high, mid, and low frequencies, respectively. The backdoor training is conducted over 1000 epochs with a learning rate of $1 \times 10^{-4}$, using the Adam optimizer, while the defensive learning is tuned in 20 epochs. All the performance is evaluated based on results from the last epoch.

\begin{table}[!ht]
    \caption{Attack performance across datasets, networks, and trigger types.}
    \centering
\fontsize{7pt}{11pt}\selectfont
\begin{tabular}{l|l|c|cc|cc|cc|cc|cc}
\hline
\centering
Dataset & Trigger & - &\multicolumn{2}{c|}{Powerline5} & \multicolumn{2}{c|}{Powerline10} & \multicolumn{2}{c|}{Powerline20} & \multicolumn{2}{c|}{Fixed} & \multicolumn{2}{c}{Random} \\
\cline{2-13}
\centering
         & Network &CA & \underline{CA} &\textbf{ ASR} & \underline{CA} &\textbf{ ASR} & \underline{CA} &\textbf{ ASR} & \underline{CA} &\textbf{ ASR}& \underline{CA} &\textbf{ ASR} \\
\hline\hline
BirdChicken & InceptionTime &90.0 &60.0 &60.0 &55.0 &90.0 &40.0 &\textbf{\textbf{100.0}} &\underline{90.0} &\textbf{\textbf{100.0}} &65.0 &\textbf{\textbf{100.0}} \\
     & LSTMFCN &90.0 &50.0 &\textbf{\textbf{100.0}} &50.0 &\textbf{\textbf{100.0}} &70.0 &90.0 &\underline{80.0} &\textbf{\textbf{100.0}} &\underline{80.0} &\textbf{\textbf{100.0}} \\
     & MACNN &85.0 &50.0 &\textbf{100.0} &50.0 &\textbf{\textbf{100.0}} &50.0 &\textbf{\textbf{100.0} }&50.0 &\textbf{\textbf{100.0}} &50.0 &\textbf{\textbf{100.0}} \\
     & TCN &80.0 &70.0 &\textbf{100.0} &70.0 &90.0 &80.0 &20.0 &\underline{85.0} &80.0 &60.0 &\textbf{\textbf{100.0}} \\
     \hline

Coffee	&InceptionTime	&100.0&60.7	&\textbf{100.0}	&53.6	&\textbf{100.0}	&71.4	&0.0	&\underline{96.4}	&\textbf{100.0}	&46.4	&\textbf{100.0} \\
&LSTMFCN	&100.0 &\underline{100.0}	&0.0	&85.7	&30.8	&82.1	&\textbf{100.0}	&\underline{100.0}	&84.6	&\underline{100.0}	&76.9\\
&MACNN	&100.0&53.6	&\textbf{100.0}	&53.6	&\textbf{100.0}	&53.6	&\textbf{100.0}	&53.6	&\textbf{100.0}	&\underline{100.0}	&\textbf{100.0}\\
&TCN	&100.0&53.6	&\textbf{100.0}	&\underline{96.4}	&0.0	&\underline{96.4}	&92.3	&\underline{96.4}	&\textbf{100.0}	&53.6	&\textbf{100.0}\\
\hline
ECG200	&InceptionTime	&87.0&73.0	&84.4	&78.0	&12.5	&37.0	&\textbf{100.0}	&\underline{83.0}	&54.7	&78.0	&96.9\\
&LSTMFCN	&87.0&66.0	&0.0	&73.0	&0.0	&75.0	&4.7	&\underline{78.0}	&76.6	&\underline{78.0}	&\textbf{92.2}\\
&MACNN	&86.0&\underline{86.0}	&3.1	&\underline{86.0}	&7.8	&\underline{86.0}	&6.2	&\underline{86.0}	&54.7	&\underline{86.0}	&31.2\\
&TCN	&89.0&83.0	&57.8	&36.0	&\textbf{100.0}	&36.0	&\textbf{100.0}	&\underline{87.0}	&\textbf{100.0}	&36.0	&\textbf{100.0}\\
\hline
ECG5000 & InceptionTime  &94.2 &\underline{94.1} &74.6 &89.9 &39.9 &72.1 &56.4 &93.9 &\textbf{100.0} &\underline{94.1} &99.8 \\
     & LSTMFCN &94.2 &90.8 & 4.8 &88.3 & 3.7 &91.7 &12.7 &\underline{94.5} &72.1 &\underline{94.5} &\textbf{98.0} \\
     & MACNN  &93.9 &58.4 &\textbf{100.0} &93.8 &72.8 &58.4 &\textbf{100.0} &\underline{88.1} &92.9 &58.4 &\textbf{100.0} \\
     & TCN  &93.3 &\underline{93.0} &34.4 &92.6 &15.4 &91.7 &55.1 &81.6 &91.5 &92.6 &72.0 \\
     \hline

Earthquakes & InceptionTime &74.1 &74.8 &\textbf{100.0} &74.8 &\textbf{100.0} &74.8 &\textbf{100.0} &74.8 &\textbf{100.0} &74.8 &\textbf{100.0} \\
     & LSTMFCN  &77.0 &74.8 &\textbf{100.0} &74.8 &\textbf{100.0} &75.5 &\textbf{100.0} &74.8 &\textbf{100.0} &74.8 &\textbf{100.0} \\
     & MACNN &74.1 &74.1 &94.3 &74.1 &91.4 &74.1 &97.1 &74.8 &\textbf{100.0} &74.8 &\textbf{100.0} \\
     & TCN &71.2 &69.1 &\textbf{100.0} &60.4 &94.3 &73.4 &91.4 &74.8 &\textbf{100.0} &74.8 &\textbf{100.0} \\
     \hline

Haptics & InceptionTime &52.3 &36.7 &75.8 &19.5 &97.2 &29.9 &86.7 &51.9 &\textbf{100.0} &\underline{52.9} &\textbf{100.0} \\
     & LSTMFCN &46.1 &41.2 & 3.2 &32.1 &17.3 &38.3 &35.5 &42.9 &\textbf{100.0} &\underline{44.5} &\textbf{100.0} \\
     & MACNN  &50.0 &\underline{19.5} &\textbf{100.0} &\underline{19.5} &\textbf{100.0} &\underline{19.5} &\textbf{100.0} &\underline{19.5} &\textbf{100.0} &\underline{19.5} &\textbf{100.0} \\
     & TCN &35.4 &\underline{36.7} &18.1 &35.4 &52.4 &34.7 &23.8 &35.1 &16.5 &36.4 &\textbf{69.8} \\
     \hline

TwoPatterns	&InceptionTime	&100.0&\underline{93.0}	&2.1	&63.2	&57.3	&88.6	&16.7	&85.3	&59.1	&63.0	&\textbf{90.8} \\
&LSTMFCN	&99.0&\underline{97.8}	&9.1	&64.7	&42.1	&93.6	&8.7	&87.6	&\textbf{67.6}	&89.8	&63.0 \\
&MACNN	&100.0&\underline{25.9}	&\textbf{100.0}	&\underline{25.9}	&\textbf{100.0}	&\underline{25.9}	&\textbf{100.0}	&\underline{25.9}	&\textbf{100.0}	&\underline{25.9}	&\textbf{100.0} \\
&TCN	&100.0&93.4	&31.4	&89.0	&25.2	&87.6	&14.0	&\underline{99.8}	&10.0	&91.2	&\textbf{51.8} \\
\hline
Wine & InceptionTime &61.1 &\underline{66.7} &\textbf{100.0} &\underline{66.7} &\textbf{100.0} &53.7 &\textbf{100.0} &\underline{66.7} &\textbf{100.0} &\underline{66.7}&\textbf{100.0} \\
     & LSTMFCN &55.6 &\underline{55.6} &\textbf{100.0} &50.0 &\textbf{100.0} &50.0 &\textbf{100.0} &50.0 &\textbf{100.0} &50.0 &\textbf{100.0} \\
     & MACNN &81.5 &\underline{81.5} &\textbf{100.0} &\underline{81.5} &\textbf{100.0} &\underline{81.5} & 0.0 &\underline{81.5} &\textbf{100.0} &\underline{81.5} &48.1 \\
     & TCN &70.4 &64.8 &\textbf{100.0} &50.0 &\textbf{100.0} &50.0 &\textbf{100.0} &\underline{75.9} &\textbf{100.0} &50.0 &\textbf{100.0} \\
     \hline
     \hline
Average & -   &81.8 & 67.1 & 67.3 &63.5 &66.9 &63.8 &66.0 &\underline{73.9} &86.3 &67.0 &\textbf{90.3}\\\hline
Count Win & -   &-  & 13 & 16 & 6 & 13 & 5 & 14 & \underline{16}& 21 &13 &\textbf{25}\\
\hline
\end{tabular}
    \label{tab:maintab}
\end{table}

\begin{table}
\centering\fontsize{7pt}{11pt}\selectfont
\caption{Results of the ablation study for \textbf{TrojanTime} (trigger type: fixed).}
\begin{tabular}{l|l|cc|cc|cc|cc}
\hline
&Dataset & \multicolumn{2}{c|}{ECG5000} & \multicolumn{2}{c|}{Coffee} & \multicolumn{2}{c|}{TwoPatterns} & \multicolumn{2}{c}{BirdChicken} \\
\hline
Models&& CA & ASR &   CA & ASR &   CA & ASR   &  CA & ASR  \\
\hline
\hline
InceptionTime	&TrojanTime	&93.9~	&100.0~	&92.9~	&100.0~	&85.3~	&59.4~	&95.0~	&100.0~ \\\cline{2-10}
&\hspace{0.5cm}w/o BN freezing 	&94.1~	&19.6$\downarrow$	&53.6$\downarrow$	&100.0~	&88.6~	&1.3$\downarrow$	&50.0$\downarrow$	&100.0~\\
&\hspace{0.5cm}w/o Logits alignment~	&90.5$\downarrow$	&100.0~	&53.6$\downarrow$	&100.0~	&67.0$\downarrow$	&64.7~	&65.0$\downarrow$	&100.0~\\
&\hspace{0.5cm}w/o $D_{\text{adv}}$	&94.1~	&43.2$\downarrow$	&96.4~	&0.0$\downarrow$	&69.1$\downarrow$	&50.4$\downarrow$	&85.0$\downarrow$ &90.0$\downarrow$\\
\hline
LSTMFCN	&TrojanTime	&93.6~	&99.9~	&46.4~	&100.0~	&88.8~	&72.7~	&90.0~	&100.0~\\\cline{2-10}

&\hspace{0.5cm}w/o BN freezing	&94.4~	&7.0$\downarrow$	&78.6~	&69.2$\downarrow$	&61.0$\downarrow$	&14.0$\downarrow$	&55.0$\downarrow$	&100.0~\\
&\hspace{0.5cm}w/o Logits alignment		&61.4$\downarrow$	&99.8~	&53.6$\downarrow$	&100.0~	&63.0$\downarrow$	&78.2~	&65.0$\downarrow$	&100.0~\\
&\hspace{0.5cm}w/o $D_{\text{adv}}$	&93.8~	&90.4$\downarrow$	&100.0~	&92.3$\downarrow$	&29.0$\downarrow$	&96.3~	&90.0~	&100.0~\\

\hline
\end{tabular}
\label{tab:ablation}
\end{table}

\subsection{Attack Performance}
Table \ref{tab:maintab} shows the comprehensive performance of the \textbf{TrojanTime} attack. The results indicate that, in most cases, the model is able to maintain its original CA while achieving a high ASR. On average, the fixed trigger performs the best, with an ASR of 86.3\%, while the CA only decreases by 7.9\% compared with the benign one. This demonstrates that our proposed method can successfully compromise a benign model without requiring access to the training data while maintaining a high CA. We also observed that the blended method (powerline) shows significantly lower ASR and CA compared to the patch masking methods (fixed, random). This is because, without carefully designed noise, the frequency distribution of the original data cannot be significantly altered, while the TSC is highly sensitive to the frequency change of the input. In contrast, masking methods can replace parts of the time series, thus obviously modifying the frequency map. The observation that high-frequency sine waves achieved higher ASR supports this conclusion. Though the core focus of this paper is not on trigger design, we still believe that with a carefully designed trigger, the ASR of the blended-type trigger could be significantly improved under the settings used in this study.


\subsection{Ablation Study}
To validate the effectiveness of each component of our proposed method, we conducted an ablation study. Table \ref{tab:ablation} shows the performance of different settings on four datasets using the InceptionTime and LSTMFCN models. If the BN layer is trainable, we can observe a general decrease in both ASR and CA. This is because BN may be influenced by the new data distribution, leading to a decrease in generalization ability. The success of the attack relies on the filters learning effective features that can strongly associate with the trigger pattern, as shown in Figure \ref{Channel Activation}.  Additionally, without logits alignment does not affect the ASR because the attack objective remains unchanged. However, the absence of original logits removes the information carried by the benign model, causing the model to forget the distribution of previous data, which is consistent with the observations in Figure \ref{TSNEa}. Furthermore, if adversarial data is not used to increase the diversity of the classes, the model may favor the most prominent class, resulting in a decrease in CA. Additionally, the lack of a proper mapping from other classes to the target class may also reduce the ASR, depending on the distribution of the injected data $D'$.

\begin{table}[h!]
\centering\fontsize{7pt}{11pt}\selectfont
\caption{Performance evaluation of defense method against \textbf{TrojanTime}.}
\begin{tabular}{l|l|l|cc|cc|cc|cc}
\hline
&& Dataset & \multicolumn{2}{c|}{ECG5000} & \multicolumn{2}{c|}{Coffee} & \multicolumn{2}{c|}{TwoPatterns} & \multicolumn{2}{c}{BirdChicken} \\
\hline
Models &Trigger Type & & CA & ASR &   CA & ASR &   CA & ASR   &  CA & ASR \\
\hline
\hline

InceptionTime	
            &Fixed	&After attack	&93.9~	&100.0~	&96.4~	&100.0~	&85.3~	&59.1~	&90.0~	&100.0~\\
                    &&After defense	&\underline{92.5}~	&18.9$\downarrow$	&46.4~	&76.9$\downarrow$	&\underline{94.3}~	&0.5$\downarrow$	&\underline{85.0}~	&50.0$\downarrow$\\
\cline{2-11}
            &Powerline5	&After attack	&94.1~	&74.6~	&60.7~	&100.0~	&93.0~	&2.1~	&60.0~	&60.0~\\
                &&After defense	&\underline{90.3}~	&41.5$\downarrow$	&\underline{89.3}~	&0.0$\downarrow$	&80.0~	&9.6~	&\underline{65.0}~	&90.0~\\
\cline{2-11}
            &Random	&After attack	&94.1~	&99.8~	&46.4~	&100.0~	&63.0~	&90.8~	&65.0~	&100.0~\\
                &&After defense	&\underline{93.0}~	&17.0$\downarrow$	&46.4~	&100.0~	&\underline{96.9}~	&4.1$\downarrow$	&50.0~	&20.0$\downarrow$\\
\hline
LSTMFCN	
        &Fixed	&After attack	&94.5~	&72.1~	&100.0~	&84.6~	&87.6~	&67.6~	&80.0~	&100.0~\\
                &&After defense	&\underline{94.0}~	&4.8$\downarrow$	&46.4~	&0.0$\downarrow$	&\underline{98.5}~	&0.2$\downarrow$	&\underline{75.0}~	&50.0$\downarrow$\\
\cline{2-11}
        &Powerline5	&After attack	&90.8~	&4.8~	&100.0~	&0.0~	&97.8~	&9.1~	&50.0~	&100.0~\\
                        &&After defense	&\underline{94.6}~	&3.4$\downarrow$	&50.0~	&0.0~	&\underline{93.4}~	&0.2$\downarrow$	&\underline{70.0}~	&90.0$\downarrow$\\
\cline{2-11}
        &Random	        &After attack	&94.5~	&98.0~	&100.0~	&76.9~	&89.8~	&63.0~	&80.0~	&100.0~\\
                        &&After defense	&\underline{94.2}~	&5.2$\downarrow$	&46.4~	&0.0$\downarrow$	&\underline{98.4}~	&0.0$\downarrow$	&{75.0}~	&20.0$\downarrow$ \\
\hline
Average         & -       &After attack	&93.6~	&74.9~	&83.9~	&76.9~	&86.1~	&48.6~	&70.8~	&93.3~\\
                        &&After defense	&\underline{93.1}~	&15.1$\downarrow$	&54.2~	&29.5$\downarrow$	&\underline{93.6}~	&2.5$\downarrow$	&\underline{70.0}~	&53.3$\downarrow$ \\
\hline
\end{tabular}
\label{tab:trigger_comparison}
\end{table}

\subsection{Defense Evaluation}
Table \ref{tab:trigger_comparison} evaluates the three most aggressive trigger types for defensive learning. It shows that our method effectively reduces the ASR across most datasets and trigger types. Some failed cases can be attributed to the insufficient number of training samples ($\sim$20), leading to almost limited bad samples that could be unlearned. From the results, we observe that our method seems to be more effective on the LSTM model compared to CNN models. One possible reason for this is that LSTMs are better suited for sequential data and are more capable of identifying patterns in temporal dependencies, making them more robust to backdoor attacks. On average, ASR is significantly decreased across all models, especially for \textit{TwoPatterns}, where CA is increased from 86.1\% to 93.6\%, almost reaching the benign model’s accuracy (100\%), and ASR reduces to 2.5\%. 
\section{Related Works}

\textbf{Backdoor Attack.} The objective of a backdoor attack is to stealthily perform a successful attack on the model, and trigger design along with poisoning methods targeting the model are key aspects of achieving this goal.  Backdoor attacks were first introduced by \cite{gu2017badnets}, which applied a fixed pixel patch in the corner of an image. To enhance stealthiness, Blended\cite{blended} was proposed, which mixes a transparent trigger with the original image. SIG\cite{SIG} stealthily embeds superimposed signals, such as ramp or sinusoidal signals, into the background of images. Dynamic\cite{Dynamic} backdoor attacks consider sample-wise triggers generated to confuse defenses and improve the chances of bypassing detection. Clean Label Attack\cite{CL} poisons the data without altering the labels. FTrojan\cite{FTrojan} focuses on applying backdoor attacks in the frequency domain. 

\noindent\textbf{Backdoor Defense.} Backdoor defenses can mainly be categorized into two types: post-process and in-process. Post-process methods typically mitigate the backdoor effect by pruning neurons or fine-tuning the model, such as FP\cite{FP}, ANP\cite{ANP}, and I-BAU\cite{I_BAU}. In-process methods, on the other hand, focus on safely training the model after the data has been poisoned. This process involves isolating or suppressing backdoor injection during training, with methods such as ABL\cite{ABL}, DBD \cite{huang2022backdoor} and DST\cite{DST}.

\section{Conclusion}
In this paper, we propose a backdoor attack method that operates in a data-inaccessible scenario. We introduce auxiliary external data and use adversarial attacks to increase its diversity. During training, we employ logits alignment and BatchNorm freezing to mitigate the concept drift problem and enhance the model’s generalization ability on clean samples. Additionally, we present a defensive method based on unlearning, where we isolate and fine-tune samples with larger responses in the rear layers. This approach successfully reduces the ASR while maintaining clean accuracy. Since this paper focuses on the design of backdoor training, we have not investigated trigger design. Therefore, in future work, we can explore the optimization of trigger design to further improve the attack’s effectiveness and stealthiness. 
\bibliographystyle{splncs04}
\bibliography{mybibliography}

\begin{thebibliography}{10}
\providecommand{\url}[1]{\texttt{#1}}
\providecommand{\urlprefix}{URL }
\providecommand{\doi}[1]{https://doi.org/#1}

\bibitem{bai2018empirical}
Bai, S., Kolter, J.Z., Koltun, V.: An empirical evaluation of generic convolutional and recurrent networks for sequence modeling. arxiv. arXiv preprint arXiv:1803.01271  \textbf{10} (2018)

\bibitem{SIG}
Barni, M., Kallas, K., Tondi, B.: A new backdoor attack in cnns by training set corruption without label poisoning. In: 2019 IEEE International Conference on Image Processing (ICIP). pp. 101--105. IEEE (2019)

\bibitem{chen2021multi}
Chen, W., Shi, K.: Multi-scale attention convolutional neural network for time series classification. Neural Networks  \textbf{136},  126--140 (2021)

\bibitem{DST}
Chen, W., Wu, B., Wang, H.: Effective backdoor defense by exploiting sensitivity of poisoned samples. Advances in Neural Information Processing Systems  \textbf{35},  9727--9737 (2022)

\bibitem{chen2021badnl}
Chen, X., Salem, A., Chen, D., Backes, M., Ma, S., Shen, Q., Wu, Z., Zhang, Y.: Badnl: Backdoor attacks against nlp models with semantic-preserving improvements. In: Proceedings of the 37th Annual Computer Security Applications Conference. pp. 554--569 (2021)

\bibitem{blended}
Chen, X., Liu, C., Li, B., Lu, K., Song, D.: Targeted backdoor attacks on deep learning systems using data poisoning. arXiv preprint arXiv:1712.05526  (2017)

\bibitem{dau2019ucr}
Dau, H.A., Bagnall, A., Kamgar, K., Yeh, C.C.M., Zhu, Y., Gharghabi, S., Ratanamahatana, C.A., Keogh, E.: The ucr time series archive. IEEE/CAA Journal of Automatica Sinica  \textbf{6}(6),  1293--1305 (2019)

\bibitem{ding2022towards}
Ding, D., Zhang, M., Huang, Y., Pan, X., Feng, F., Jiang, E., Yang, M.: Towards backdoor attack on deep learning based time series classification. In: 2022 IEEE 38th International Conference on Data Engineering (ICDE). pp. 1274--1287. IEEE (2022)

\bibitem{dong2024boosting}
Dong, C., Li, Z., Zheng, L., Chen, W., Zhang, W.E.: Boosting certified robustness for time series classification with efficient self-ensemble. arXiv preprint arXiv:2409.02802  (2024)

\bibitem{dong2023swap}
Dong, C.G., Zheng, L.N., Chen, W., Zhang, W.E., Yue, L.: Swap: Exploiting second-ranked logits for adversarial attacks on time series. In: 2023 IEEE International Conference on Knowledge Graph (ICKG). pp. 117--125. IEEE (2023)

\bibitem{goodfellow2014explaining}
Goodfellow, I.J., Shlens, J., Szegedy, C.: Explaining and harnessing adversarial examples. arXiv preprint arXiv:1412.6572  (2014)

\bibitem{gu2017badnets}
Gu, T., Dolan-Gavitt, B., Garg, S.: Badnets: Identifying vulnerabilities in the machine learning model supply chain. arXiv preprint arXiv:1708.06733  (2017)

\bibitem{huang2022backdoor}
Huang, K., Li, Y., Wu, B., Qin, Z., Ren, K.: Backdoor defense via decoupling the training process. arXiv preprint arXiv:2202.03423  (2022)

\bibitem{ismail2020inceptiontime}
Ismail~Fawaz, H., Lucas, B., Forestier, G., Pelletier, C., Schmidt, D.F., Weber, J., Webb, G.I., Idoumghar, L., Muller, P.A., Petitjean, F.: Inceptiontime: Finding alexnet for time series classification. Data Mining and Knowledge Discovery  \textbf{34}(6),  1936--1962 (2020)

\bibitem{jiang2023backdoor}
Jiang, Y., Ma, X., Erfani, S.M., Bailey, J.: Backdoor attacks on time series: A generative approach. In: 2023 IEEE Conference on Secure and Trustworthy Machine Learning (SaTML). pp. 392--403. IEEE (2023)

\bibitem{karim2017lstm}
Karim, F., Majumdar, S., Darabi, H., Chen, S.: Lstm fully convolutional networks for time series classification. IEEE access  \textbf{6},  1662--1669 (2017)

\bibitem{ABL}
Li, Y., Lyu, X., Koren, N., Lyu, L., Li, B., Ma, X.: Anti-backdoor learning: Training clean models on poisoned data. Advances in Neural Information Processing Systems  \textbf{34},  14900--14912 (2021)

\bibitem{liang2024enhancing}
Liang, W., Li, Z., Chen, W.: Enhancing financial market predictions: Causality-driven feature selection. arXiv preprint arXiv:2408.01005  (2024)

\bibitem{lin2024unveiling}
Lin, W., Liu, L., Wei, S., Li, J., Xiong, H.: Unveiling and mitigating backdoor vulnerabilities based on unlearning weight changes and backdoor activeness. arXiv preprint arXiv:2405.20291  (2024)

\bibitem{FP}
Liu, K., Dolan-Gavitt, B., Garg, S.: Fine-pruning: Defending against backdooring attacks on deep neural networks. In: International symposium on research in attacks, intrusions, and defenses. pp. 273--294. Springer (2018)

\bibitem{liu2018trojaning}
Liu, Y., Ma, S., Aafer, Y., Lee, W.C., Zhai, J., Wang, W., Zhang, X.: Trojaning attack on neural networks. In: 25th Annual Network And Distributed System Security Symposium (NDSS 2018). Internet Soc (2018)

\bibitem{Dynamic}
Nguyen, T.A., Tran, A.: Input-aware dynamic backdoor attack. Advances in Neural Information Processing Systems  \textbf{33},  3454--3464 (2020)

\bibitem{pelletier2019temporal}
Pelletier, C., Webb, G.I., Petitjean, F.: Temporal convolutional neural network for the classification of satellite image time series. Remote Sensing  \textbf{11}(5), ~523 (2019)

\bibitem{CL}
Shafahi, A., Huang, W.R., Najibi, M., Suciu, O., Studer, C., Dumitras, T., Goldstein, T.: Poison frogs! targeted clean-label poisoning attacks on neural networks. Advances in neural information processing systems  \textbf{31} (2018)

\bibitem{shen2022leads}
Shen, S., Chen, W., Xu, M.: What leads to arrhythmia: Active causal representation learning of ecg classification. In: Australasian Joint Conference on Artificial Intelligence. pp. 501--515. Springer (2022)

\bibitem{FTrojan}
Wang, T., Yao, Y., Xu, F., An, S., Tong, H., Wang, T.: An invisible black-box backdoor attack through frequency domain. In: European Conference on Computer Vision. pp. 396--413. Springer (2022)

\bibitem{wu2021adversarial}
Wu, D., Wang, Y.: Adversarial neuron pruning purifies backdoored deep models. Advances in Neural Information Processing Systems  \textbf{34},  16913--16925 (2021)

\bibitem{ANP}
Wu, D., Wang, Y.: Adversarial neuron pruning purifies backdoored deep models. Advances in Neural Information Processing Systems  \textbf{34},  16913--16925 (2021)

\bibitem{yang2021rethinking}
Yang, W., Lin, Y., Li, P., Zhou, J., Sun, X.: Rethinking stealthiness of backdoor attack against nlp models. In: Proceedings of the 59th Annual Meeting of the Association for Computational Linguistics and the 11th International Joint Conference on Natural Language Processing (Volume 1: Long Papers). pp. 5543--5557 (2021)

\bibitem{I_BAU}
Zeng, Y., Chen, S., Park, W., Mao, Z.M., Jin, M., Jia, R.: Adversarial unlearning of backdoors via implicit hypergradient. arXiv preprint arXiv:2110.03735  (2021)

\end{thebibliography}
\end{document}